\documentclass[12pt]{article}
\usepackage{amsmath, amsthm, amssymb,graphicx}
\begin{document}
\title{\textbf{Scale invariant correlations\\and the distribution of prime numbers}} 
\author{B.~Holdom%
\thanks{bob.holdom@utoronto.ca}\\
\emph{\small Department of Physics, University of Toronto}\\[-1ex]
\emph{\small Toronto ON Canada M5S1A7}}
\date{}
\maketitle
\begin{abstract}
Negative correlations in the distribution of prime numbers are found to display a scale invariance. This occurs in conjunction with a nonstationary behavior. We compare the prime number series to a type of fractional Brownian motion which incorporates both the scale invariance and the nonstationary behavior. Interesting discrepancies remain. The scale invariance also appears to imply the Riemann hypothesis and we study the use of the former as a test of the latter.
\end{abstract}

\section{Scale Invariance}
The distribution of prime numbers has been a source of fascination for mathematicians for centuries. Of interest to physicists is the fact that the distribution of prime numbers exhibits certain characteristics that are usually associated with chaotic and complex systems. The data on the prime number series is plentiful and a surprisingly rich structure is uncovered when this data is probed in various ways \cite{n2}. The closely associated zeros of the Riemann zeta function have been studied even more intensely from a physics viewpoint \cite{n14}.

We shall show that prime number differences display a scale invariance in their correlations. The differences also display a nonstationary behavior that coexists with the scale invariance. We shall identify a random process, fractional Brownian motion of a lesser known type, that displays these and other properties of the prime number series. But there are still quantitative differences which suggest that a further generalization of fractional Brownian motion remains to be developed.  The resolution of this puzzle could have practical value for descriptions of complex systems in physics and economics. In addition we shall describe how a breakdown of the Riemann hypothesis would not be compatible with the preservation of scale invariance. This provides some further insight and an additional test of the Riemann hypothesis in terms of concepts familiar to physicists.

Consider the difference
\begin{equation}
{\rm Li}(x)-\pi(x).
\label{e4}\end{equation}
${\rm Li}(x)$ is the logarithmic integral function, the principal value of $\int_0^x 1/\ln(t) dt$, and $\pi(x)$ is the prime counting function, the number of primes less than or equal to $x$. The Riemann hypothesis is equivalent an upper bound on the absolute value of this difference, with a particular example being $|{\rm Li}(x)-\pi(x)|<\sqrt{x}\ln(x)/(8\pi)$ for $x\ge2657$ \cite{n1}. We extract from (\ref{e4}) the discrete series
\begin{equation}
b(i)\equiv{\rm Li}(p_i)-i,
\end{equation}
where the $i$th prime number is $p_i$. The positive increments ${\rm Li}(p_{i+1})-{\rm Li}(p_i)$ have a mean close to unity and a distribution close to exponential for sufficiently large sample size. Thus to first approximation $b(i)$ executes something close to a random walk as it ``diffuses'' away from zero for increasing $i$. But it is certainly not a true random walk \cite{n3}. The Riemann hypothesis is a nontrivial constraint on how far $|b(i)|$ can depart from zero, while at the same time $b(i)$ remains positive to some extremely large value of $i$.

A study of correlations among the differences, $b(i+j)-b(i)$, will help to characterize more precisely how $b(i)$ differs from a true random walk. For convenience the correlations we consider will be constructed from the sequences $\{b(1\cdot2^n), b(2\cdot2^n), b(3\cdot2^n), ... ,b(2^m\cdot2^n)\}$ determined by integers $m\ge 3$ and $n\ge 0$. From these we obtain the sequences of the $2^m-1$ successive differences, 
\begin{equation}
\Delta(m,n)=\{b(2\cdot2^n)-b(1\cdot2^n), b(3\cdot2^n)-b(2\cdot2^n), ...\}
.\end{equation}
We wish to investigate the correlations among the entries of $\Delta(m,n)$. For increasing $n$ these sequences will probe correlations on larger and larger scales, and $n$ can be thought to label a scale transformation from the nearest neighbor differences ($n=0$) to larger differences. For each $\Delta(m,n)$ consider the two sequences $x$ and $y$ formed by dropping the first $k$ and last $k$ elements of $\Delta(m,n)$ respectively. Then calculate the Pearson correlation 
\begin{equation}
C_k(m,n)=\frac{\sum_i(x_i-\overline{x})(y_i-\overline{y})}{\sqrt{\sum_i(x_i-\overline{x})^2}\sqrt{\sum_i(y_i-\overline{y})^2}}
.\end{equation}
In the language of discrete time series, $C_k(m,n)$ is an autocorrelation of lag $k$ calculated from a sample of $2^m$ primes.

We highlight the sequences for $m$ from 10 to 16 with $n$ ranging from 0 up to $37-m$. We don't need all primes up to the $2^{37}$th prime, which is about $3.8\times10^{12}$, but only those to fill out the $\Delta(m,n)$ sequences. The results for the correlations $C_1(m,n)$ and $C_2(m,n)$ are shown in Fig.~(1) and Fig.~(2) respectively. Note that we have added a 0.1 vertical spacing between the points of different $m$.
\begin{center}\includegraphics[scale=0.5]{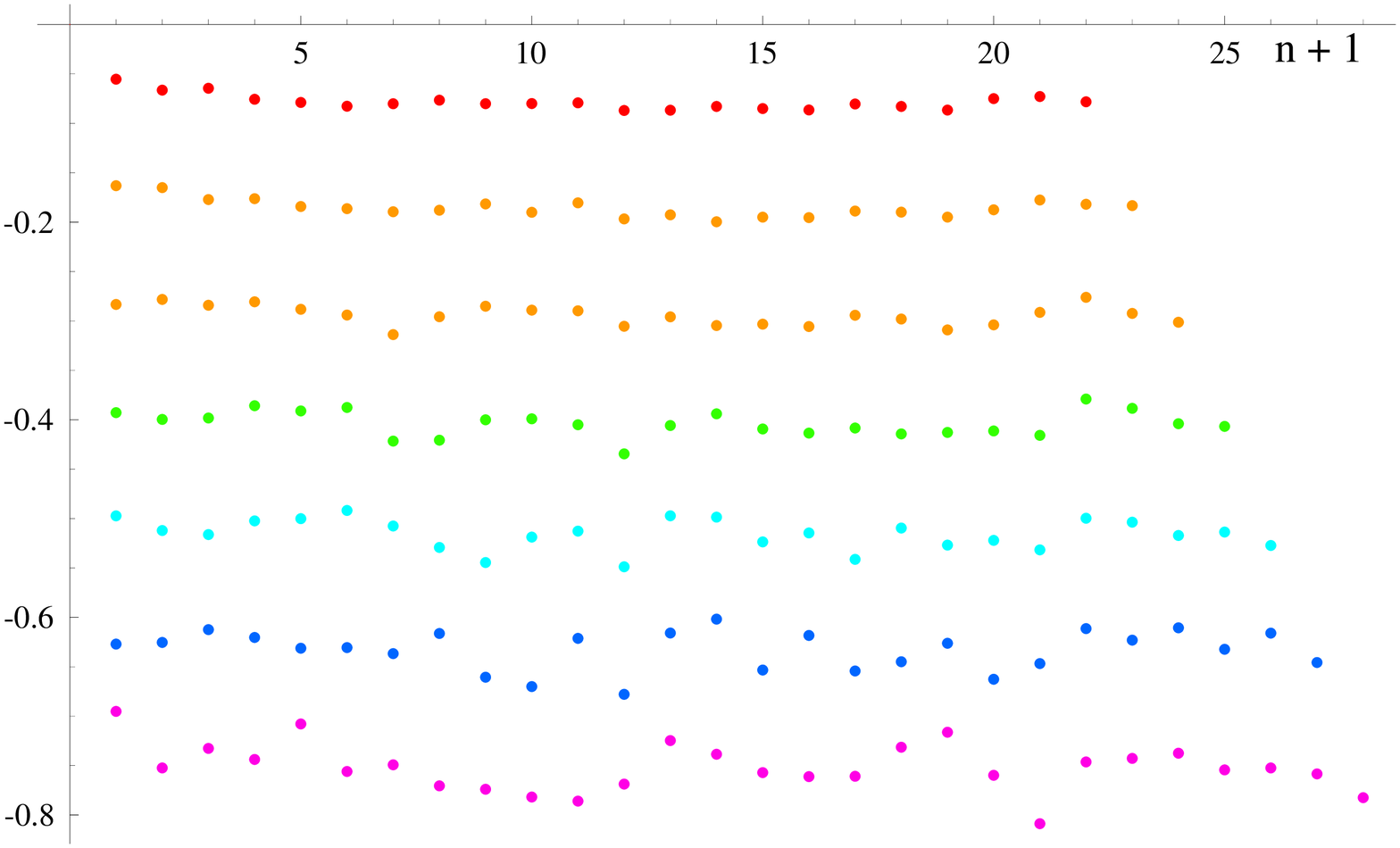}\end{center}
\vspace{-1ex}\noindent Figure 1: $C_1(m,n)-(16-m)/10$ for $m=16, 15, 14, 13, 12, 11, 10$ from top to bottom.
\begin{center}\includegraphics[scale=0.5]{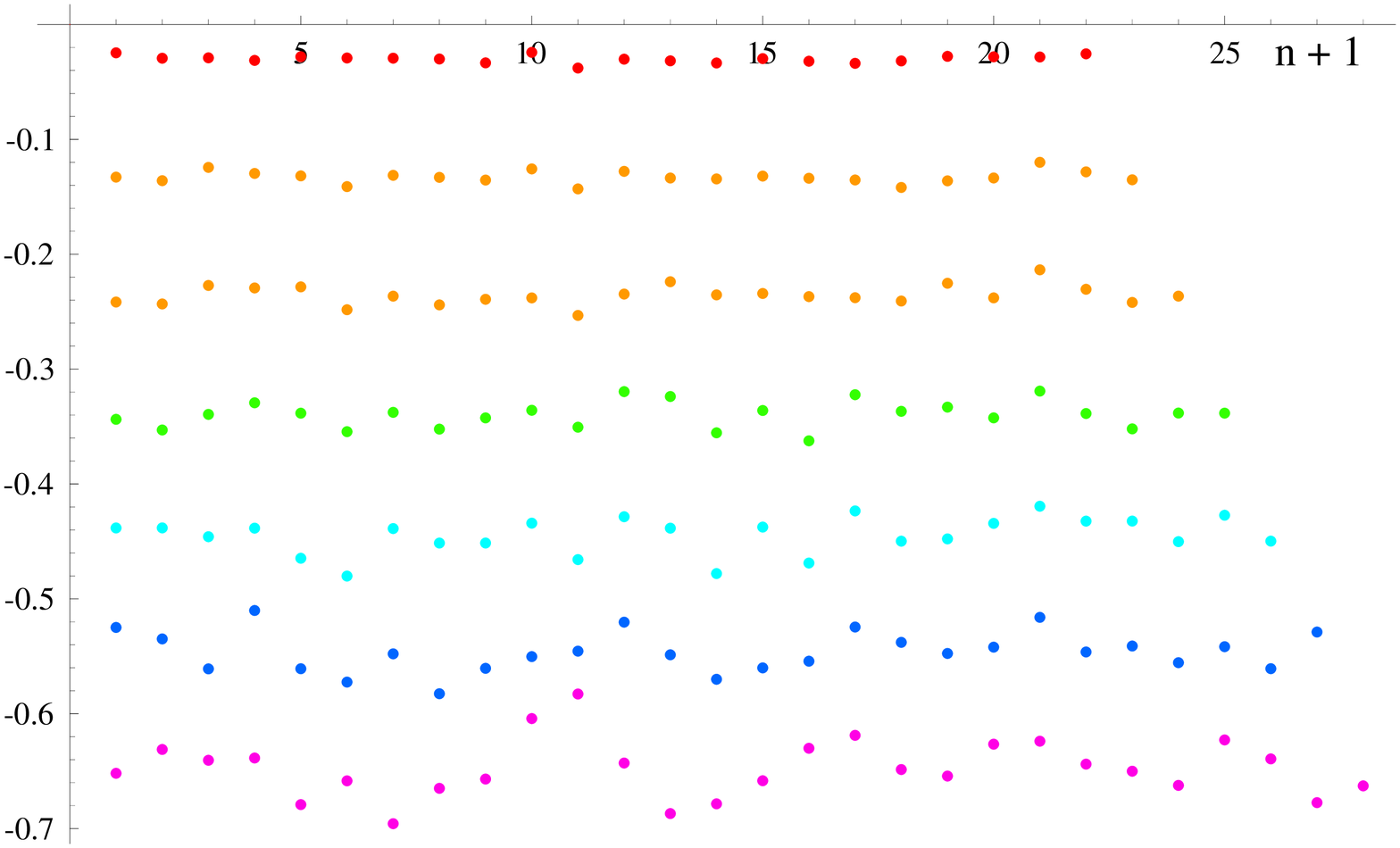}\end{center}
\vspace{-1ex}\noindent Figure 2: $C_2(m,n)-(16-m)/10$ for $m=16, 15, 14, 13, 12, 11, 10$ from top to bottom.
\vspace{2ex}

We perform a least-squares linear fit to the data for each $m$ and obtain the 95\% confidence interval for the slopes, as listed in the following Table.
\begin{center}\begin{tabular}{|c|c|c|c|c|}\hline $m$ & Slope($C_1$) & $C_1(m)$ & Slope($C_2$) & $C_2(m)$\\\hline 16& [$-0.0011$, $-0.00015$]  & $-0.078\pm0.0079$ & [$-0.00026$, $0.00019$] & $-0.030\pm0.003$ \\\hline 15& [$-0.0012$, $-0.00012$]  & $-0.085\pm0.0094$ & [$-0.00036$, $0.00037$] & $-0.033\pm0.005$ \\\hline  14& [$-0.0011$, 0.00005]  & $-0.094\pm0.010$ & [$-0.00025$, $0.00078$] & $-0.036\pm0.008$ \\\hline 13& [$-0.0010$, 0.0005]  & $-0.104\pm0.013$ & [$-0.00039$, $0.00094$] & $-0.040\pm0.012$ \\\hline 12& [$-0.0013$, 0.0004]  & $-0.116\pm0.015$ & [$-0.00032$, $0.0014$] & $-0.045\pm0.016$ \\\hline 11 & [$-0.0011$, 0.0010] & $-0.133\pm0.020$ & [$-0.0007$, $0.0011$] & $-0.046\pm0.018$ \\\hline 10 & [$-0.0020$, 0.0004] & $-0.152\pm0.024$ & [$-0.0011$, $0.0013$] & $-0.048\pm0.025$ \\\hline\hline 9& [$-0.002$, $0.001$]  & $-0.17\pm0.04$ & [$-0.001$, $0.002$] & $-0.06\pm0.04$ \\\hline  8& [$-0.004$, 0.001]  & $-0.20\pm0.06$ & [$-0.002$, $0.003$] & $-0.06\pm0.07$ \\\hline 7& [$-0.003$, 0.002]  & $-0.24\pm0.07$ & [$-0.002$, $0.003$] & $-0.10\pm0.08$ \\\hline 6& [$-0.03$, 0.003]  & $-0.31\pm0.09$ & [$-0.006$, $0.003$] & $-0.09\pm0.13$ \\\hline 5 & [$-0.006$, 0.003] & $-0.37\pm0.15$ & [$-0.0007$, $0.0011$] & $-0.09\pm0.22$ \\\hline 4 & [$-0.006$, 0.008] & $-0.45\pm0.20$ & [$-0.0011$, $0.0013$] & $-0.09\pm0.35$\\\hline 3 & [$-0.02$, 0.004] & $-0.49\pm0.27$ & [$-0.002$, $0.01$] & $0.05\pm0.45$ \\\hline \end{tabular}\end{center}\vspace{2ex}
For completeness we have extended the Table to smaller values of $m$; for $m$ from 3 to 9, $n$ ranges from 0 to $40-m$. All slopes are consistent with zero. Therefore we define $C_k(m)$ as the mean of the $C_k(m,n)$'s for each $m$. Note that the quoted errors on the $C_k(m)$'s are the standard deviations ($\approx\frac{1}{\sqrt{2^m}}$) of the $C_k(m,n)$'s about their means; the standard deviations of the $C_k(m)$'s themselves are $\frac{1}{\sqrt{38-m}}$ times smaller.

The first point is that these correlations, which would be consistent with zero for a true random walk, are instead nonvanishing and negative. And although the $|C_k(m)|$ decrease for increasing $m$, the significance of their departure from zero increases. Negative correlations indicate a tendency for mean reversion. This phenomenon is clearly connected with the Riemann hypothesis, since it indicates that the near random movements have less tendency to drive $b(i)$ from zero than would otherwise be the case for a true random walk. The second point is that the correlations for each $m$ fluctuate around a mean value as the scale parameter $n$ is changed. The correlations thus display a (statistical) self-similarity or scale invariance.

In Figs.~(3) and (4) we display how the variance and kurtosis of the distribution of differences depend on the scaling parameter $n$. The variance for a true random walk would grow as $\sigma^2 2^n$ where $\sigma^2$ is the variance of the nearest neighbor differences, and this corresponds to the horizontal line in Fig.~(3). As expected the observed variances grow as a smaller power. From the behavior of the kurtosis we see that the distribution of differences quickly approaches the Gaussian form for increasing $n$. As mentioned above, the distribution of the nearest neighbor differences ($n=0$) has the form of a shifted exponential. At small $n$ one can see that there are also corresponding small anomalies in the correlations and variances.
\begin{center}\includegraphics[scale=0.5]{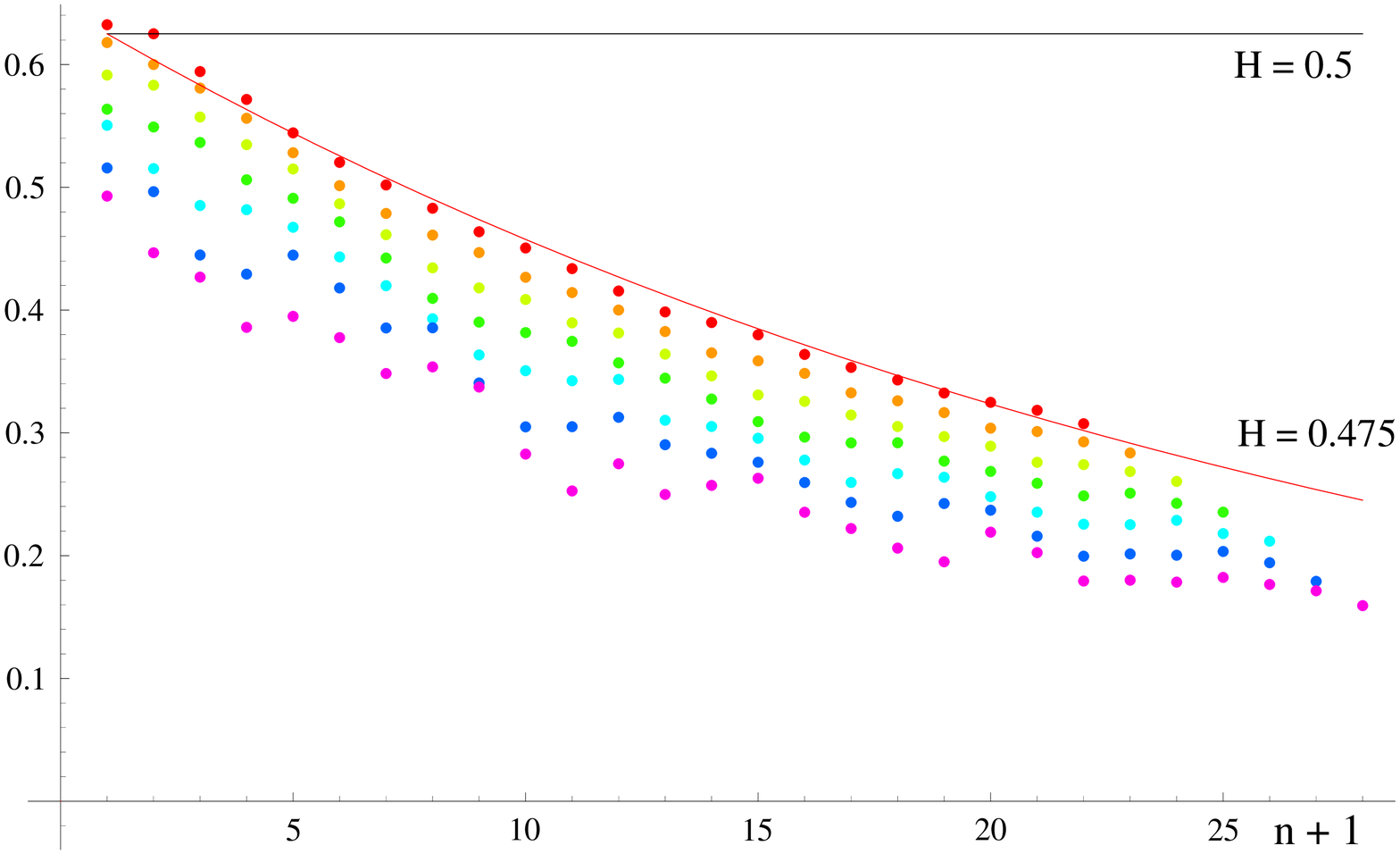}\end{center}
\vspace{-1ex}\noindent Figure 3: Variance$[\Delta(m,n)]/2^n$ for $m=16, 15, ... 10$ from top to bottom. The lines are examples of $\sigma^2 2^{(2H-1)n}$.
\vspace{2ex}
\begin{center}\includegraphics[scale=0.5]{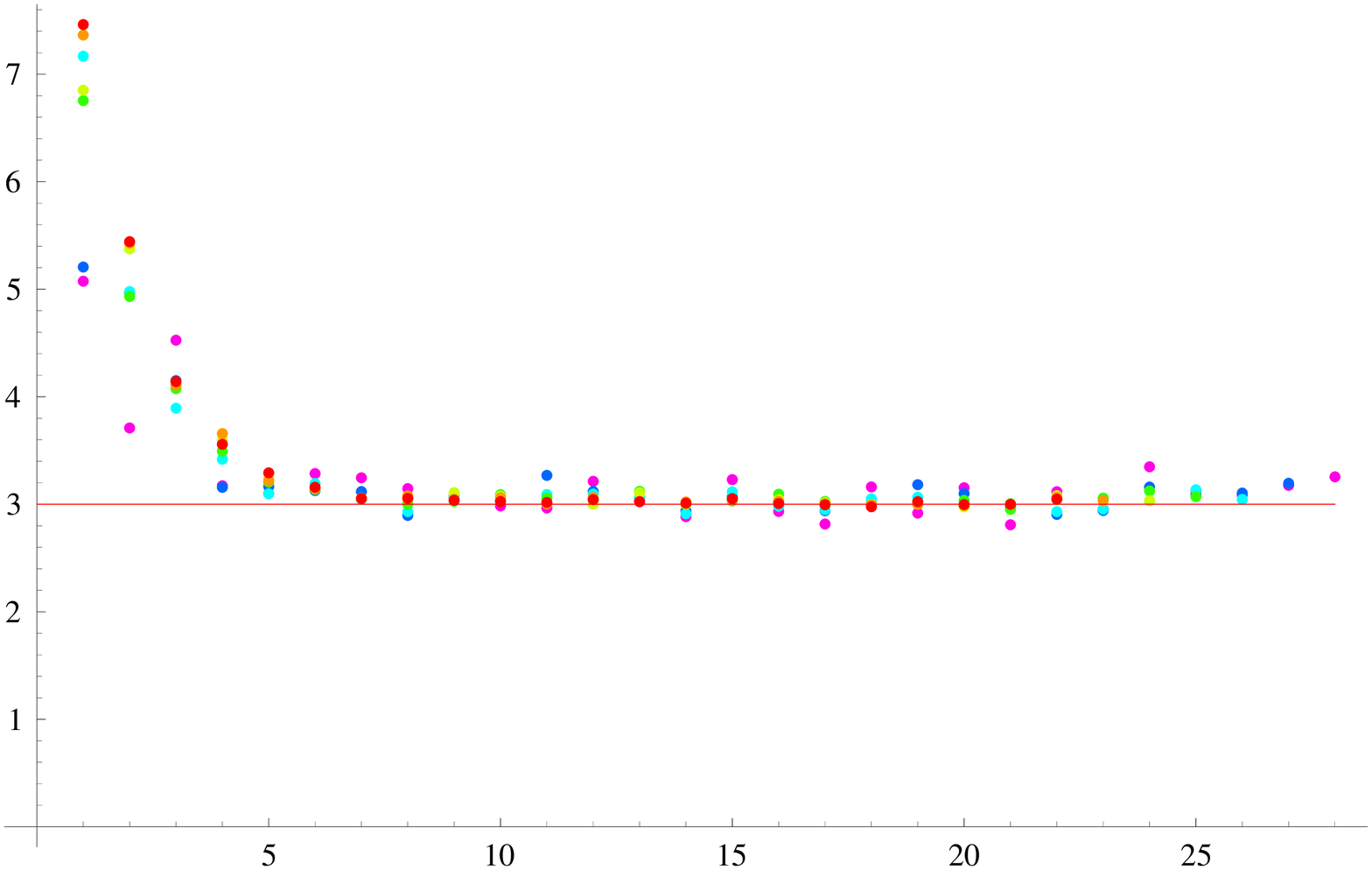}\end{center}
\vspace{-1ex}\noindent Figure 4: Kurtosis$[\Delta(m,n)]$, defined such that the Gaussian value is 3. The higher $m$ values approach this line more closely.
\vspace{2ex}

\section{Fractional Brownian motion}
Fractional Brownian motion (fBm) \cite{n4} is a random Gaussian process $B_H(t)$ with expectations $E[B_H(t)]=0$ and $E[B_H(t)^2]=\sigma^2|t|^{2H}$. The Hurst exponent is $0<H<1$. Let us consider a discrete series of values $B_H(i)$ for integers $i$. If we define the differences $\Delta_H(i,j)=B_H(i+j)-B_H(i)$ for nonvanishing integers $j$ then the correlations for lag $k$ are
\begin{equation}
C_H(k,j,i)=\frac{E[\Delta_H(i,j)\Delta_H(i+k,j)]}{\sqrt{E[\Delta_H(i,j)^2]E[\Delta_H(i+k,j)^2]}}
.\label{e5}\end{equation}
The scale invariance (self-similarity) of fBm corresponds to $C_H(\lambda k,\lambda j,\lambda i)=C_H(k,j,i)$. This is the hallmark of fBm, and it is the desired property to mirror the scale invariance ($n$ independence) in the prime number correlations.

If $B_H(i)$ is defined to have stationary differences then the correlations are independent of $i$ and are uniquely determined \cite{n4},
\begin{equation}
C^{\rm I}_H(k,j)=\frac{|k+j|^{2H}-2|k|^{2H}+|k-j|^{2H}}{2|j|^{2H}}
.\label{e1}\end{equation}
To obtain the desired negative correlations we must choose $0<H<1/2$. Then all $C^{\rm I}_H(k,1)=C^{\rm I}_H(-k,1)<0$ except $C^{\rm I}_H(0,1)=1$, and we also note the sum rule $\sum_{k=-\infty}^\infty C^{\rm I}_H(k,1)=0$. $C^{\rm I}_H(1,1)$ and $C^{\rm I}_H(2,1)$ are displayed in Fig.~(5) below. Mean reversion is characterized by the fact that if given only two values $B_H(i_1)$ and $B_H(i_2)$, the expectation for $B_H(\infty)$ is $(B_H(i_1)+B_H(i_2))/2$ for any $0<H<1/2$. Ordinary Brownian motion has $H=1/2$ and its expectation for $B_H(\infty)$ is $B_H(\max(i_1,i_2))$.

The fBm process as defined above misses an essential feature. The correlations (\ref{e1}) are independent of $i$ and so they do not capture the observed $m$ dependence, $C_k(m)$, of the prime number correlations. For example by comparing the $C_1(m)$'s in the Table above to $C^{\rm I}_H(1,1)$, the implied values of $H$ are $(0.441, 0.436, 0.428, 0.421, 0.411, 0.398, 0.381)$ for $m=16, 15, ...10$.  This $m$ dependence represents a nonstationary behavior of the increments of the prime number series.  There is also $m$ dependence in the variances in Fig.~(3), and for $m=16$ the implied value of $H$ is $\approx0.475$. This extraction of $H$ from the scale dependence of the variances is similar in spirit to a rescaled range analysis. The latter was applied \cite{n5} to the distribution of large primes (the 25th million and 50th million primes) where $H\approx0.46$ was obtained. But such an analysis does not in itself disentangle the scale invariance from the nonstationary behavior. As with the variances there is also a strong sensitivity, for example, to a simple re-weighting $b(i)\rightarrow w(i)b(i)$. The correlations display much more robustly the scale invariance and the nonstationary behavior.

The nonstationarity (a breaking of translation invariance) is perhaps not surprising given that the prime number series has a beginning. In fact there is another type of fractional Brownian motion that better reflects these properties. It was actually considered first \cite{n11,n10,n4} due to its very simple definition in terms of a stochastic integral,
\begin{equation}
B_H(t)=\sqrt{2H}\int_0^t(t-u)^{H-1/2}dB(u),\quad t\ge0
,\end{equation}
where $B(u)$ represents ordinary Brownian motion. With this definition $B_H(t)$ has a beginning as well as nonstationary increments. This has been referred to as Type II fBm in \cite{n12}, where a comparison with the better known Type I fBm is made. An early study \cite{n10} of the former was concerned with $H=1$, but the covariance obtained there can be generalized to other $H$ by use of a hypergeometric function,
\begin{equation}
E[B_H(t)B_H(s)]=\frac{2H}{H+1/2}t^{H+1/2}s^{H-1/2}{}_2F_1(1,\frac{1}{2}-H,\frac{3}{2}+H,\frac{t}{s}),\quad s>t.
\label{e6}\end{equation}
When $(s-t)/t$ is small this approaches the corresponding covariance for Type I fBm, which is $(|t|^{2H}+|s|^{2H}-|s-t|^{2H})/2$.

The correlations as defined in (\ref{e5}) can be obtained for type II fBm by using (\ref{e6}); they now have $i$ dependence and they can be denoted by $C^{\rm II}_H(k,j,i)$. The scale invariance remains intact, $C^{\rm II}_H(\lambda k,\lambda j,\lambda i)=C^{\rm II}_H(k,j,i)$. In addition the correlations of type I fBm can be obtained as a limit, $\lim_{i\rightarrow\infty}C^{\rm II}_H(k,j,i)=C^{\rm I}_H(k,j)$. Away from this limit $|C^{\rm II}_H(k,j,i)|>|C^{\rm I}_H(k,j)|$ with the largest difference at $i=0$. In Fig.~(5) we compare $C^{\rm II}_H(k,1,0)$ and $C^{\rm I}_H(k,1)$ as a function of $H$ for $k=1,2$.
\begin{center}\includegraphics[scale=0.5]{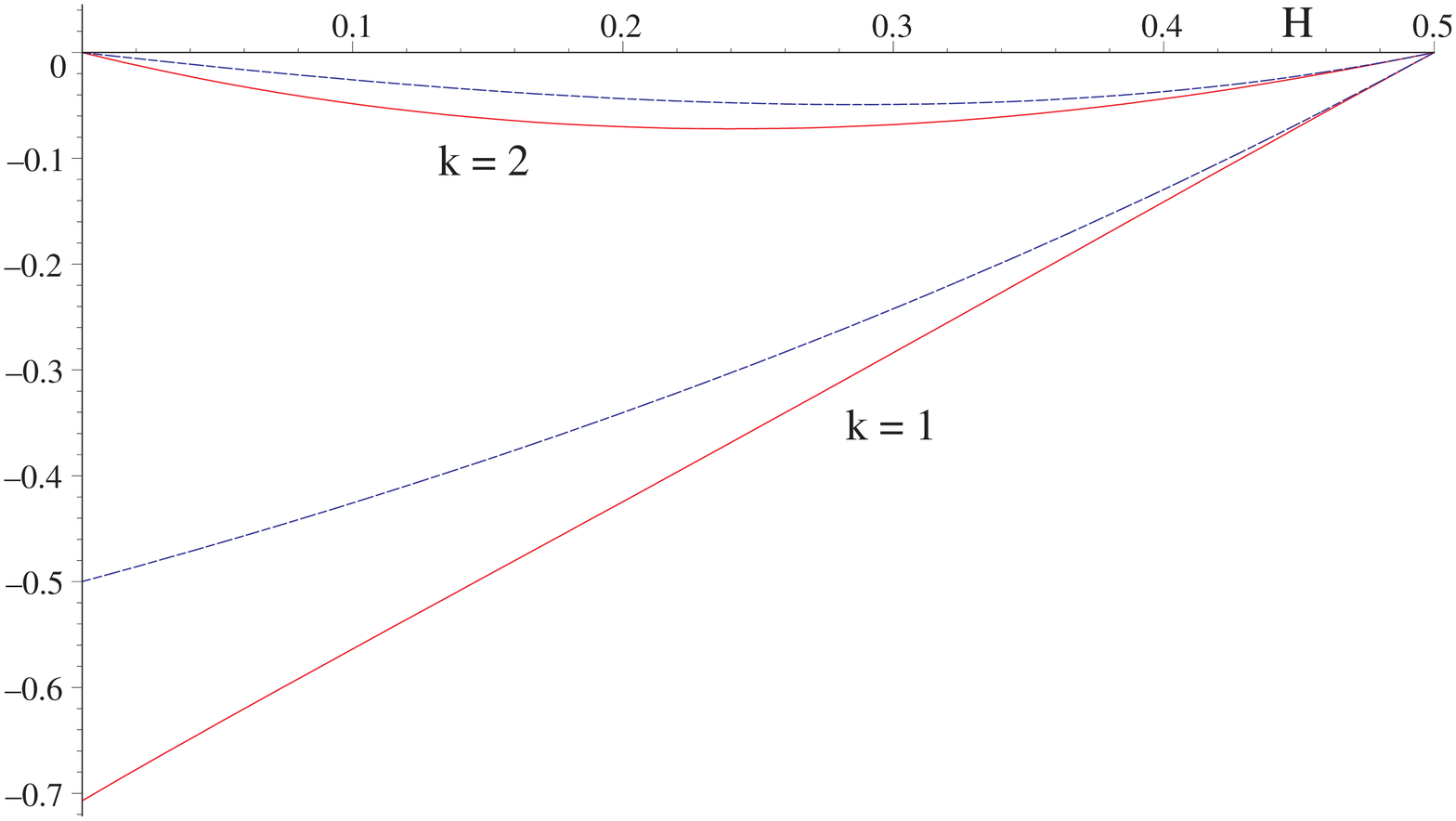}\end{center}
\vspace{-1ex}\noindent Figure 5: $C^{\rm I}_H(k,1)$ (dashed blue) and $C^{\rm II}_H(k,1,0)$ (solid red) as a function of $H$ for $k=1,2$.
\vspace{2ex}

The correlations among prime numbers $|C_k(m)|$ decrease with increasing $m$, and this is nicely mirrored by the similar behavior of $C^{\rm II}_H(k,1,i)$ with increasing $i$. (Note that our calculation of $C_k(m)$ would correspond to an average of $C^{\rm II}_H(k,1,i)$ over a range of $i$ that grows with $m$.)  Thus the properties of Type II fBm agree at least qualitatively with the prime number series. We display in Fig.~(6) the values of $C_k(m)$ for a range of $k$ and compare to the values spanned by $C^{\rm II}_H(k,1,\infty)$ and $C^{\rm II}_H(k,1,0)$ for $H=0.4$. The overall $k$ dependence is in good qualitative agreement. But we see that the variability of $C^{\rm II}_H(k,1,i)$ with $i$ is significantly less than the variability of $C_k(m)$ with $m$. Thus while the increase in $|C^{\rm II}_H(k,1,0)|$ relative to $|C^{\rm I}_H(k,1)|$ goes in the right direction it does not go nearly far enough.
\begin{center}\includegraphics[scale=0.5]{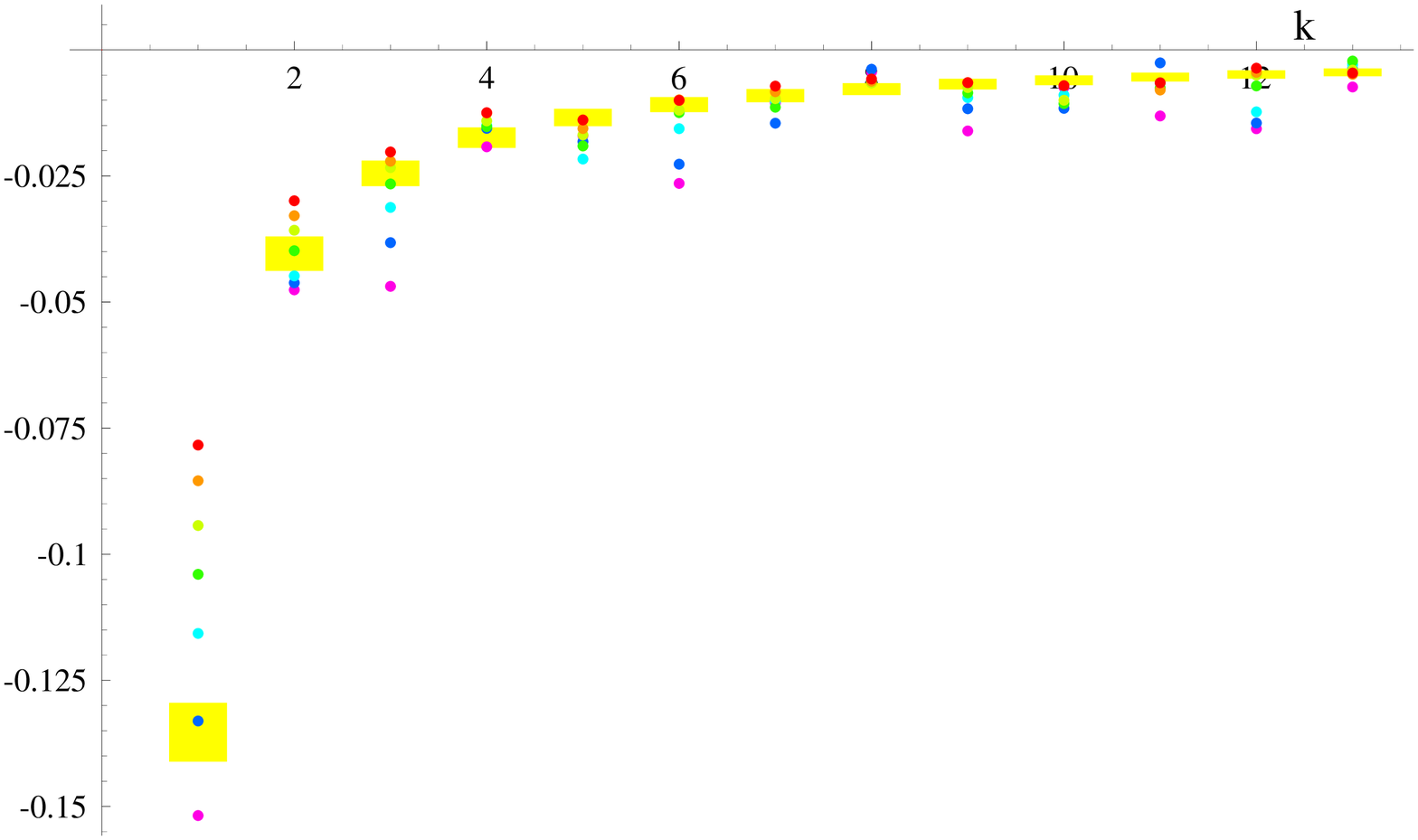}\end{center}
\vspace{-1ex}\noindent Figure 6: $C_k(m)$ for $m=16, 15, ... 10$ from top to bottom. The upper and lower bounds on the yellow rectangles are $C^{\rm I}_H(k,1)=C^{\rm II}_H(k,1,\infty)$ and $C^{\rm II}_H(k,1,0)$ respectively for $H=0.4$.
\vspace{2ex}

We note that the implied $H$ from the correlations increases with increasing $m$, while the implied $H$ from the variances, if anything, decreases (from Fig.~(3)). Thus for sufficiently large $m$ these implied values of $H$ may come together and perhaps the type II fBm process is more closely approached. But for the values of $m$ we have studied it is clear that nonstationary behavior is more pronounced for the prime number $b(i)$ series than it is for the type II fBm process. This raises the question of whether there is a generalization of a fBm process in which the nonstationarity can be adjusted while leaving the scale invariance intact. There are various multifractal generalizations of fBm, such as ``multifractional Browning motion'' \cite{n8} and ``fractional Brownian motion in multifractal time'' \cite{n9}. For these processes the probability distribution of the increments may no longer be Gaussian and its shape can change as a function of the scaling parameter. Clearly Fig.~(4) showing the kurtosis does not see evidence away from small $n$ for this type of multifractal behavior. In addition these generalizations tend to lose the global scale invariance, in which case they are not appropriate for the $b(i)$ series.

\section{The Riemann hypothesis}
Here we focus on the scale invariance and its relation to the Riemann hypothesis. Assuming that this hypothesis is true then there is a relation between $b(i)$ and the nontrivial zeros of the zeta function $\zeta(s)$,\footnote{For (\ref{e2}) to become an equality the ``1'' should be replaced by a slowly varying function of $p_i$ that approaches 1 from above as $p_i$ increases.}
\begin{equation}
\frac{b(i)}{\sqrt{p_i}/\log(p_i)}\approx 1+2\sum_{\gamma>0} \frac{\sin(\gamma\log(p_i))}{\gamma}
,\label{e2}\end{equation}
where these zeros are of the form $\frac{1}{2}+i\gamma$. The more common version \cite{n3} of (\ref{e2}) has $p_i$ replaced by $x$ and $b(i)$ replaced by the difference in (\ref{e4}), ${\rm Li}(x)-\pi(x)$.\footnote{The re-weighted version of the $b(i)$ series appearing on the LHS of (\ref{e2}) is central to other investigations \cite{n15}, and one may wonder whether this re-weighted series could be used in place of $b(i)$ in the previous analysis. The scale invariance of the correlations would still persist, but the increments of the re-weighted series would have variances far from the constant $\sigma^2$ of fBm. }

Now consider the case where the Riemann hypothesis is violated, which corresponds to zeros appearing off the $\frac{1}{2}+i t$ line and in the strip between $i t$ and $1+i t$. Such zeros must be symmetrically placed about the real axis and the $\frac{1}{2}+i t$ line. The result of such a quartet of zeros at a distance $d$ from the $\frac{1}{2}+i t$ line and a distance $T$ from the real axis is to add the following term (ignoring sub-leading terms) to the RHS of (\ref{e2}),
\begin{equation}
2\frac{p_i^d}{T}\sin(T\log(p_i))
.\label{e3}\end{equation}
If such a quartet of zeros existed, then for large enough $p_i$ the new term in (\ref{e3}) will eventually dominate the sum. This is well known as is the fact that there are no such zeros up to some very large value of $T$, currently of order $2.4\times10^{12}$ \cite{n7}.

Nevertheless we would like to see how the scale invariance would be affected by such zeros. We therefore study the modified series,
\begin{equation}
b(i)\rightarrow \left[\frac{b(i)}{\sqrt{p_i}/\log(p_i)}+2\frac{p_i^d}{T}\sin(T\log(p_i))\right]\frac{\sqrt{p_i}}{\log(p_i)}
.\end{equation}
Figs.~(7) and (8) display the modified correlations and variances for $d=.25$ and $T=20000$. A dramatic violation of scale invariance has occurred.\footnote{The kurtosis also deviates from its Gaussian value.}

The violation causes a bifurcation in the correlations for some value of $m$; around this value the correlations at sufficiently large $n$ are pushed away from their original values in opposite directions.  By varying $T$ we find that the value of $m$ near the bifurcation is such that $2^m\approx T$. This is an interesting connection between the number of primes in the samples where a bifurcation is detected, and the location of a $d>0$ zero of the zeta function. The value of $d$ affects the value of $n$ where the bifurcation becomes visible; as $d$ is decreased the departures from scale invariance move to larger $n$ values. But for a fixed $d$ and $T$ there are $m$ and $n$ sufficiently large such that the correlations tend to unity.
\begin{center}\includegraphics[scale=0.5]{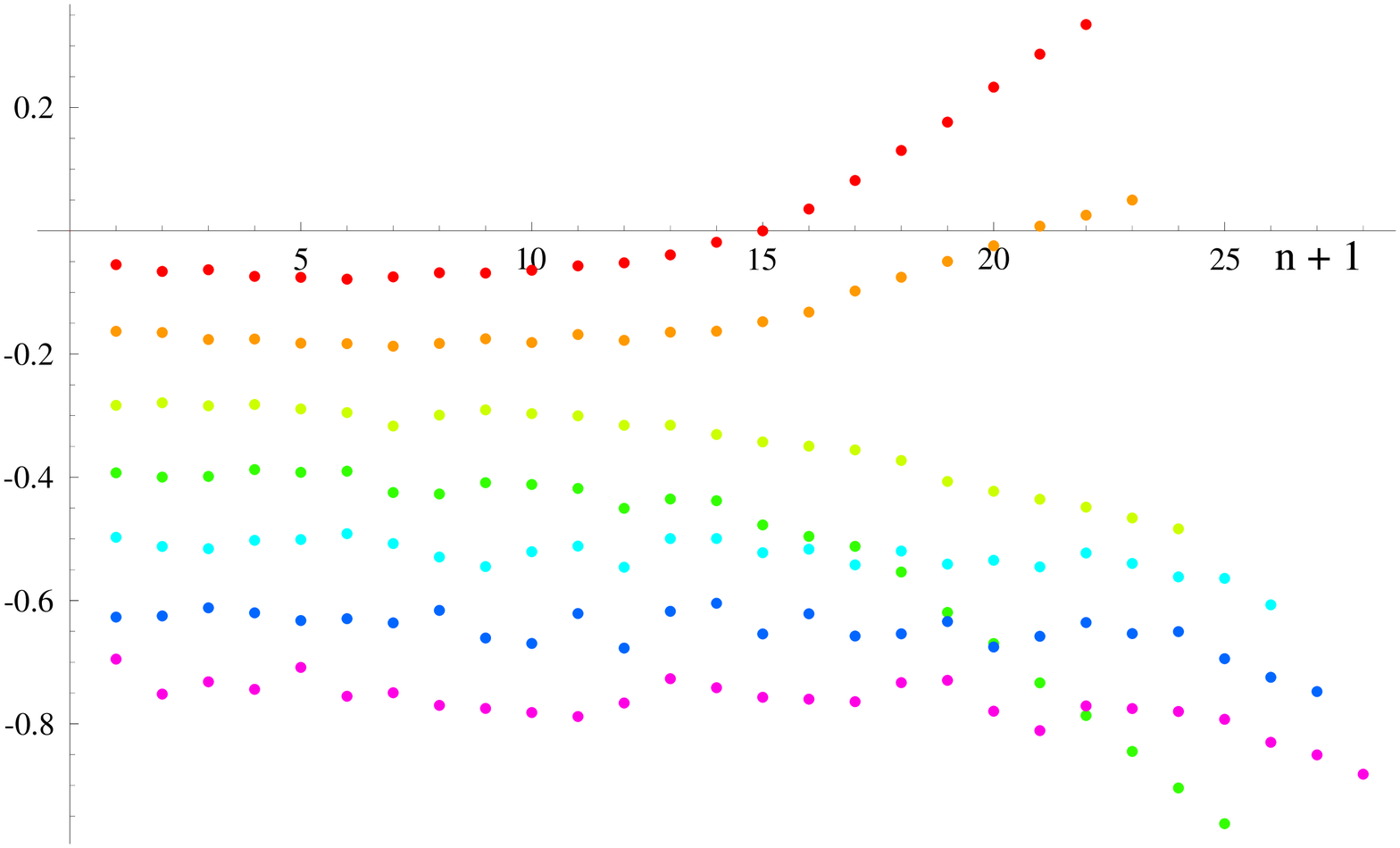}\end{center}
\vspace{-1ex}\noindent Figure 7: $C_1(m,n)-(16-m)/10$ for $m=16, 15, ...10$ from top to bottom, showing a violation of the Riemann Hypothesis.
\begin{center}\includegraphics[scale=0.5]{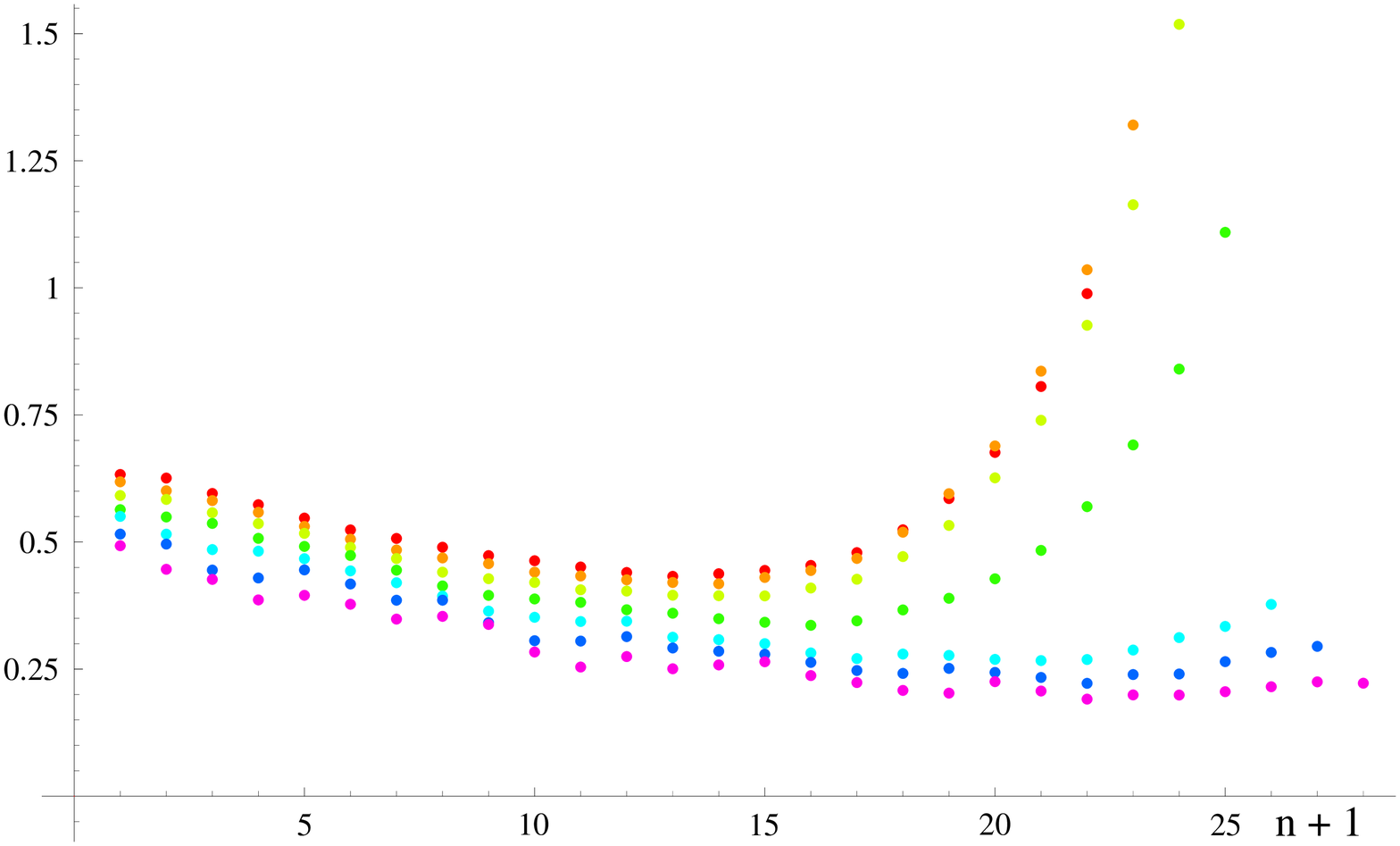}\end{center}
\vspace{-1ex}\noindent Figure 8: Variance$[\Delta(m,n)]/2^n$ for $m=16, 15, ... 10$ from top to bottom, showing a violation of the Riemann Hypothesis.
\vspace{2ex}

We may estimate more generally the primes that would be needed to detect the broken scale invariance characterized by some zero with $d>0$ at height $T$.  The value of $n$ needed to easily detect the distortion in the correlations is $n\approx 1/d^2$.  Thus $n2^m\approx T/d^2$ is the number of primes needed. The largest prime $p_i$ of this set would have $i=2^n2^m\approx2^\frac{1}{d^2}T$. We note that $p_i$ as a function of $i$ has been calculated up to about $10^{22}$ \cite{n3}. If we use the corresponding $i$ as an upper bound then this limits the sensitivity to a violation of the Riemann hypothesis. For example for $d=.25$ it would be possible to rule out such zeros with $T$ as large as $10^{17}$. 

We have shown that the existence of a single quartet of zeros that violate the Riemann hypothesis would show up as a breakdown of the scale invariance.  There could be a whole spectrum, possibly infinite, of zeros away from the $\frac{1}{2}+i t$ line. Although we do not prove it here, it appears that any such violation of the Riemann hypothesis would modify the correlations and/or the variances such that the scale invariance evident in Figs.~(1-3) would not persist to arbitrarily high $m$ and $n$. Conversely, the statement that scale invariance does persist to arbitrarily large $m$ and $n$ is a statement that is at least as strong as the Riemann hypothesis. If the preservation of scale invariance is indeed a stronger statement, then it would not only enforce all nontrivial zeros to lie on the line $\frac{1}{2}+it$, but it would also constrain the distribution of zeros along that line. Precisely what form this constraint takes is an open question.

The distribution of zeros of the zeta function, and in particular the probability distribution of the spacings, has been a topic of intense study (e.g.~see \cite{n7}). The most naive possibility, that there exists a scale invariance of correlations among zeros as we have found for the primes,  does not appear to be the case. Correlations of nearest neighbor spacings at lag $k$ and at high $t$ were explored in \cite{n13} where it was found that $C^{\rm zeros}_k\approx-0.32/k^2$. This behavior is quite different from the behavior for fBm, as given in (\ref{e1}), which thus suggests that the correlations are not scale invariant. A Hurst exponent was extracted \cite{n6} using a rescaled range analysis where it was found that $H\approx0.1$ characterizes the distribution of zeros at high $t$. This scaling did not apply for zeros at low $t$ and even for high $t$ the scaling fit was poor. As mentioned below (\ref{e2}) the full $x$ dependence of the function ${\rm Li}(x)-\pi(x)$ is reproduced from the knowledge of the zeros, whereas the series $b(i)$ only samples this function at a discrete set of values.  Thus while the zeros appear to contain much more information, much of it may be superfluous as far as the scale invariance is concerned.

In summary we have described the scale invariance of the negative correlations that exist in the distribution of prime numbers. The scale invariance ($n$ independence) was separated from a nonstationarity ($m$ dependence) and these results encompass both large and small primes. The prime numbers are providing a nontrivial and fundamental manifestation of scale invariance. We have described a lesser known version of fractional Brownian motion that enjoys a simple definition and that displays the same properties as the prime number series. But to obtain precise agreement, the prime numbers are suggesting some further scale invariant generalization of fBm. We have also argued that this structure is of mathematical interest, since the scale invariance appears to imply the Riemann hypothesis. There may thus be motivation to directly test the scale invariance to much larger primes, and we estimated the number of primes needed to set competitive limits on violations of the Riemann hypothesis.

\section*{Acknowledgments}
This work was supported in part by the Natural Science and Engineering Research Council of Canada.


\begin{thebibliography}{11}
\bibitem{n2} M.~Wolf, Applications of statistical mechanics in number theory, Physica A274 (1999) 149, and references therein.
\bibitem{n14} M.~V.~Berry and J.~P.~Keating, $H = xp$ and the Riemann zeros, Supersymmetry and Trace Formulae: Chaos and Disorder, edited by I.~V.~Lerner et.~al., New York: Plenum  (1999) 355, and references therein,\\\verb$http://www.phy.bris.ac.uk/people/berry_mv/the_papers/Berry306.pdf$.
\bibitem{n1} L.~Schoenfeld, Sharper bounds for the Chebyshev functions $\theta (x)$ and $\psi (x)$. II. Math. Comp. 30 (1976), no. 134, 337.
\bibitem{n3} A.~Granville, Analytic number theory, The Princeton Companion to Mathematics, Princeton University Press (2008), also available at \verb$http://www.dms.umontreal.ca/~andrew/$.
\bibitem{n4} B.~Mandelbrot and J.~van Ness, Fractional Brownian motions, fractional noises and applications, SIAM Review 10(4) (1968) 422.
\bibitem{n5} O.~Shanker, Correlations in prime number distribution and L-function zeros, \verb$http://sites.google.com/site/primenumbergaps/Home$.
\bibitem{n11} P.~Levy, Random functions: general theory with special reference in Laplacian random functions, 
Univ.~California Publ.~Statist.~1 (1953) 331.
\bibitem{n10} J.~A.~Barnes and D.~W.~Allan, A statistical model of flicker noise, Proc. of the IEEE 54 (1966) 176.
\bibitem{n12} D.~Marinucci, P.~M.~Robinson, Alternative forms of fractional Brownian motion, Journal of Statistical Planning and Inference 80 (1999) 111.
\bibitem{n8} R.~F.~Peltier and J.~Levy Vehel, Multifractional Brownian motion: definition and preliminary results, Rapport de recherche de lÕINRIA, no. 2645 (1995),\\\verb$http://hal.inria.fr/docs/00/07/40/45/PDF/RR-2645.pdf$
\bibitem{n9} B.~Mandelbrot, Fractals and scaling in finance. Springer New York, 1997.
\bibitem{n15} M.~Rubinstein and P.~Sarnak, Exp.~Math.~3 (1994) 173; A.~Wintner, Amer.~J.~Math.~63 (1941) 233.
\bibitem{n7} X.~Gourdon, The $10^{13}$ first zeros of the Riemann zeta function, and zeros computation at very large height, 2004,\\\verb$http://numbers.computation.free.fr/Constants/Miscellaneous/zetazeros1e13-1e24.pdf$.
\bibitem{n13} A. M. Odlyzko. The $10^{20}$-th zero of the Riemann zeta function and 175 million of its neighbors, 1992, \verb$http://www.dtc.umn.edu/~odlyzko/unpublished/index.html$.
\bibitem{n6} O.~Shanker, Random matrices, generalized zeta functions and self-similarity of zero distributions, J. Phys. A: Math. Gen. 39 (2006) 13983.

\end{thebibliography}
\end{document}